\documentclass[aps,prl,reprint,floatfix,longbibliography,twocolumn]{revtex4-2}
\usepackage{amsmath,amssymb,amsthm,hyperref,graphicx}
\hypersetup{hidelinks}
\hypersetup{colorlinks=true, linkcolor=blue, citecolor=red, urlcolor=blue}

\begin{document}

\title{Breaking surface plasmon excitation constraint via surface spin waves}
\author{H. Y. Yuan$^{1,2}$}
\email{hyyuan@zju.edu.cn}
\author{Yaroslav M. Blanter$^{2}$}
\affiliation{$^{1}$Institute for Advanced Study in Physics, Zhejiang University, 310027 Hangzhou, China}
\affiliation{$^{2}$Department of Quantum Nanoscience, Kavli Institute of Nanoscience, Delft University of Technology, 2628 CJ Delft, The Netherlands}

\date{\today}

\begin{abstract}
Surface plasmons in two-dimensional (2D) electron systems have attracted great attention for their promising light-matter applications. However, the excitation of a surface plasmon, in particular, transverse-electric (TE) surface plasmon, remains an outstanding challenge due to the difficulty to conserve energy and momentum simultaneously in the normal 2D materials. Here we show that the TE surface plasmons ranging from gigahertz to terahertz regime can be effectively excited and manipulated in a hybrid dielectric, 2D material and magnet structure. The essential physics is that the surface spin wave supplements an additional freedom of surface plasmon excitation and thus greatly enhances the electric field in the 2D medium. Based on widely-used magnetic materials like yttrium iron garnet (YIG) and manganese difluoride ($\mathrm{MnF}_2$), we further show that the plasmon excitation manifests itself as a measurable dip in the reflection spectrum of the hybrid system while the dip position and the dip depth can be well controlled by an electric gating on the 2D layer and an external magnetic field. Our findings should bridge the fields of low-dimensional physics, plasmonics and spintronics and open a novel route to integrate plasmonic and spintronic devices.
\end{abstract}

\maketitle
{\it Introduction.---} Plasmons are collective excitations of electronic charge density in metallic structures. In three-dimensional (3D) systems, one has to overcome a gap of several electronvolts to excite the bulk plasma oscillations, which makes it challenging to be manipulated. The situation in two-dimensional (2D) systems is very different since the plasmon frequency is usually proportional to $\sqrt{q}$ with $q$ being the propagating wavevector of plasmons \cite{ZayatsPR2005}, implying that the excitation energy can be desirably tuned far below the optical regime. Another benefit of a 2D configuration is the electrical tunability of the Fermi energy and thus of the charge carrier density \cite{RodinNRP2020,Ukhtary2020}. As a result, surface plasmons in 2D materials, for example, graphene, have attracted significant attention with the well-developed fabrication technology of 2D materials \cite{RodrigoSci2015,IranzoSci2018,EpsteinSci2020,RodinNRP2020,Ukhtary2020,ZhaoNature2023}. In particular, transverse magnetic (TM) plasmons are broadly studied while transverse electric (TE) plasmons are seldom studied for its restrictive excitation condition. For usual 2D systems with the parabolic electron dispersion, it is widely believed that the TE plasmons are not present for their positive imaginary component of conductivity, which is well described by the Drude model \cite{JablanPRB2009}. For graphene, it was theoretically proposed that the sign of the imaginary part of the conductivity may reverse near the spectral onset of intraband scattering to unlock the TE modes \cite{MikhailovPRL2007}. However, the resulting TE plasmons locating in infra and terahertz regime are yet to be verified.

On the other hand, spin waves -- collective excitations of spins in ordered magnets -- can carry information even in magnetic insulators, which largely reduces the Joule heating problem during information processing \cite{ChumakNP2015,PirroNRM2021}. Spintronic systems are also easily integrated with other physical systems, for example, photons, qubits and phonons, to form hybrid systems for multifunctional information processing \cite{YuanReview2022,RameshtiPR2022}. The frequency of spin waves ranges from gigahertz (GHz) in ferromagnets (FM) to terahertz (THz) in antiferromagnets (AFM) \cite{BaltzRMP2018}. This makes it possible to couple them to surface plasmons in 2D materials that have a continuous spectrum \cite{DyrdalPRB2023,GhoshPRB2023,CostaNL2023}. Hybrid 2D materials with magnetic films, which stimulated a lot of interest recently \cite{MendesPRL2015,SongSci2018,TakiguchiSci2019,GhiasiNN2021,WangCP2021,HuJAC2022,WangSR2023}, also provide an accessible platform to investigate the hybrid magnon-plasmon excitation.

\begin{figure}
	\centering
	\includegraphics[width=0.45\textwidth]{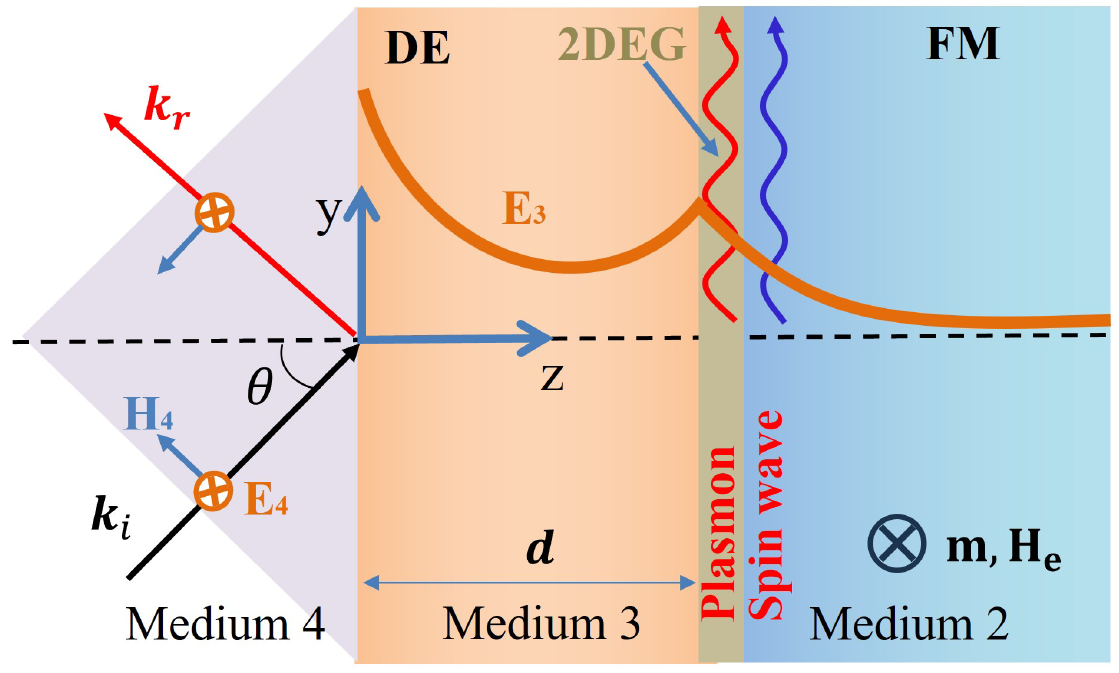}\\
	\caption{Schematic of the modified Otto configuration composing of a prism, a dielectric layer, a 2DEG layer and a ferromagnetic layer. An incident electromagnetic wave induces an evanescent wave in the dielectric layer above a critical incident angle. The evanescent wave propagates toward $+e_z$ direction and excites surface plasmons in the 2DEG layer as well as surface spin waves in the magnetic layer.}\label{fig1}
\end{figure}

In this \textit{Letter}, we show how the conventional constraint on a TE surface plasmon excitation can be overcome by the interplay of surface plasmons and spin waves. In particular, we investigate the wave propagation in a hybrid dielectric (DE), 2D electron gas (2DEG) and magnetic insulator structure as shown in Fig. \ref{fig1}. An incident electromagnetic wave first induces a surface wave at the interface of media 4-3 and an evanescent wave inside medium 3. The evanescent wave propagates towards the 2DEG layer and excites TE surface plasmons and surface spin waves simultaneously free from the constraint of 2DEG conductivity. The frequency of this surface plasmon-magnon polariton can be well tuned by an external magnetic field and falls into the GHz regime for FMs and THz for AFMs. For comparison, plasmons are barely generated when a magnet is replaced by a (non-magnetic) dielectric material. Furthermore, the excitation of the surface plasmon manifests as a sharp and robust dip in the reflection spectrum of the layered structure, which is feasible to be observed in experiments. These findings give a state-of-art demonstration of surface-plasmon excitations in hybrid 2D material-magnet structures and they should open a promising avenue to study the interplay of spintronics, nanophotonics and low-dimensional physics. 

{\it Physical model and the excitation spectrum.}---Let us first look at the excitation spectrum of the hybrid system DE/2DEG/FM(DE) shown in Fig. \ref{fig1}, where the DE and FM layers are semi-infinite. The electromagnetic properties of the hybrid structure should satisfy the Maxwell's equations
\begin{equation}\label{MaxwellDBHE}
\nabla \times \mathbf{E} = -\partial_t \mathbf{B},~ \nabla \times \mathbf{H} = \partial_t \mathbf{D},
\end{equation}
where $\mathbf{E}$ and $\mathbf{H}$ are respectively electric and magnetic fields, while $\mathbf{D}=\epsilon_0\epsilon \mathbf{E}$ and $\mathbf{B}=\mu_0 (\mathbf{H}+\mathbf{M})$ are respectively the electric displacement and the magnetic inductance with $\epsilon_0$, $\mu_0$ and $\epsilon$ being the vacuum permittivity, vacuum permeability and material permittivity, respectively. After eliminating the electric components, Eqs. \eqref{MaxwellDBHE} can be combined to
\begin{equation}\label{MaxwellHM}
(\nabla^2 +k^2) \mathbf{H}- \nabla(\nabla \cdot \mathbf{H}) + k^2\mathbf{M} =0,
\end{equation}
where $k^2 = \epsilon \mu_0 \omega^2$, $\mathbf{M}=M_s\mathbf{m}$ with $M_s$ being the saturation magnetization and $\mathbf{m}$ the normalized magnetization vector of the FM layer.

On the other hand, the magnetization dynamics in the FM layer is governed by the Landau-Lifshitz-Gilbert (LLG) equation \cite{Landau1935,Gilbert2004,TserkovnyakRMP2005}
\begin{equation}\label{LLG}
	\partial_t \mathbf{m} =-\gamma \mathbf{m} \times \mathbf{H}_{\mathrm{eff}} + \alpha \mathbf{m} \times \partial_t \mathbf{m}.
\end{equation}
The first and second terms on the right-hand side of Eq. \eqref{LLG} respectively describe the precessional and damped motion of the magnetization toward the effective field $\mathbf{H}_\mathrm{eff}$ with $\gamma$ and $\alpha$ being the gyromagnetic ratio and the Gilbert damping parameter. In general, $\mathbf{H}_\mathrm{eff}$ is a sum of the external field $\mathbf{H}_e$, the dipolar field $\mathbf{H}$, the crystalline anisotropy field, and the exchange field. It is assumed that the external field is applied along the $x$-axis $\mathbf{H}_e=H_ee_x$ and is strong enough to generate a uniform equilibrium state $\mathbf{M}_0=M_se_x$. Then the spin-wave excitation above this ground state can be represented as $\mathbf{M}=\mathbf{M}_0 + M_ye_y + M_ze_z$ with $M_{y,z} \ll M_s$ and the dynamics of $(M_y,M_z)$ is derived by linearizing the LLG equation \eqref{LLG} around $\mathbf{M}_0$ as
\begin{equation}\label{llgmymz}
\left (\begin{array}{c}
  M_y \\
  M_z
\end{array} \right )=
\left ( \begin{array}{cc}
  \kappa & -i\nu\\
  i \nu & \kappa
\end{array} \right )
\left (\begin{array}{c}
  H_y \\
  H_z
\end{array} \right ),
\end{equation}
where $\kappa=(\omega_h-i\alpha \omega) \omega_m/((\omega_h-i\alpha \omega)^2 - \omega^2)$, $\nu=\omega_m \omega/((\omega_h-i\alpha \omega)^2 - \omega^2)$ with $\omega_h=\gamma H_e$, $\omega_m = \gamma M_s$. This linearization procedure is sufficient to describe low-energy excitation of spin waves \cite{KittelPR1948,KefferPR1952}. Without loss of generality, we have neglected the exchange field, because it does not contribute significantly to the low-energy excitation in the soft magnets like yttrium iron garnet (YIG).

By substituting Eq. \eqref{llgmymz} into the Maxwell's equations \eqref{MaxwellHM}, we can derive self-contained equations of $H_y$ and $H_z$.
We consider an incident wave with momentum $\mathbf{k}^{(i)} = (0,k_4 \cos \theta, k_4 \sin \theta)$. Then the spins mainly oscillate in the $y$ and $z$ directions, and the combined LLG and Maxwell equations in medium 2 read
\begin{equation}\label{llgmaxwell}
\left ( \begin{array}{cc}
  \partial_{zz} + k_2^2(1+\kappa) & -\partial_{yz} -i k_2^2 \nu \\
  -\partial_{yz} + i k_2^2 \nu &  \partial_{yy} + k_2^2(1+\kappa)
\end{array} \right )
\left (\begin{array}{c}
  H_y \\
  H_z
\end{array} \right )=0.
\end{equation}
By solving Eqs \eqref{llgmaxwell}, we derive the surface spin-wave mode with $\mathbf{H}_2 = (0,H_{2,y}^{(-)}, H_{2,z}^{(-)})e^{ik_{2,y}y-\kappa_2z},$
$\mathbf{E}_2=(E_{2,x}^{(-)}, 0,0)e^{ik_{2,y}y-\kappa_2z}$, $H_{2,y}^{(-)}=i\kappa_2E_{2,x}^{(-)}/(\mu_0 \omega)$ and $k_{2,y}, \kappa_2, \omega$ are related to each other by the determinantal equation. Unless stated otherwise, we always label the wavevector and decay exponent in medium $i$ by $k_{i,y}$ and $\kappa_i$, and they satisfy $k_{i,y}^2 - \kappa_i^2=\omega^2/c^2 \epsilon_i$ ($i=3,4$). The upper indices $(\pm)$ indicate the exponential increase/decay modes in the $+z$-axis. Note that such surface modes can be interpreted as Damon-Eshbach modes beyond the magnetostatic limit \cite{DamonPR1960,DamonJPCS1961}.

In the dielectric medium 3, $\mathbf{M}=0$, the TE wave solution to the Maxwell's equations reads $\mathbf{H}_3 = (0,H_{3,y}^{(+)}, H_{3,z}^{(+)})e^{ik_{3,y}y+\kappa_3z},$
$\mathbf{E}_3=(E_{3,x}^{(+)}, 0,0)e^{ik_{3,y}y+\kappa_3z}$ with the relation $H_{3,y}^{(+)} = -i\kappa_3 E_{3,x}^{(+)}/(\mu_0 \omega)$. At the interface of media 3 and 2, the tangential components of electric field should be continuous while the tangential components of magnetic field are connected by the surface electric currents $j_{x}=\sigma E_{2,x}^{(-)}$ corresponding to surface plasmons excitations in the 2DEG layer, {\em i.e.}
\begin{equation}\label{dispersion_boundarycondition}
H_{3,y}^{(+)}-H_{2,y}^{(-)}=\sigma E_{2,x}^{(-)},~E_{3,x}^{(+)}=E_{2,x}^{(-)},
\end{equation}
where we shifted the $z=0$ plane to the interface of 2DEG for simplicity and imposed the requirement of in-plane momentum conservation $k_{3,y}=k_{2,y}\equiv q$. The semiclassical approximation for plasmons as surface electric currents associated with electromagnetic wave emission is a standard approach \cite{MikhailovPRL2007,JablanPRB2009}. Nontrivial solutions of Eqs. \eqref{dispersion_boundarycondition} exist provided
\begin{equation}\label{resonance_condition}
\sqrt{q^2-\frac{\omega^2 \epsilon_3}{c^2}} + \frac{\epsilon_2\omega^2}{c^2\delta_p} -i \mu_0 \omega \sigma =0,
\end{equation}
where $\delta_p$ ($p\in\{ \mathrm{FM},\mathrm{DE}\}$) depends on the nature of medium 2 such that
\begin{subequations}
\begin{align}
\delta_\mathrm{FM}&=-i\left ( k_{2,y}\frac{k_{2,z}^2-k_2^2 (1+ \kappa)}{k_{2,y}k_{2,z} -iv k_2^2} - k_{2,z} \right), \\
\delta_\mathrm{DE} &= k_2^2/\kappa_2.
\end{align}
\end{subequations}
This is our first key result.
When medium 2 is a dielectric, the resonance condition is reduced to the familiar form in literature by inserting $\delta_\mathrm{DE}$ into Eq. \eqref{resonance_condition} \cite{MikhailovPRL2007}. This condition only has real solutions when $\sigma$ is purely imaginary and otherwise has a negative imaginary component. Therefore, it cannot be fulfilled for conventional 2DEG whose conductivity is well described by the Drude model \cite{Jackson1998EM,MikhailovPRL2007,ChangNC2016} i.e. $\sigma = \sigma_0 E_F/(\pi\Gamma - i\pi\hbar \omega)$ with $\sigma_0= e^2/4\hbar$, $E_F$ being the Fermi energy and $\Gamma$ being the relaxation rate of carriers. This implies that the TE surface plasmons cannot be resonantly excited using the conventional Otto setup.
\begin{figure}
	\centering
	\includegraphics[width=0.49\textwidth]{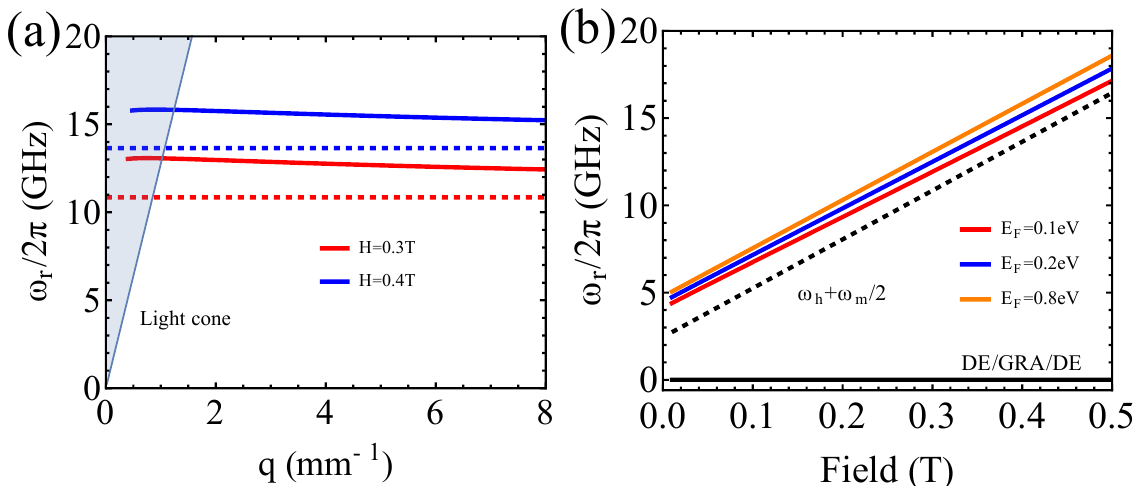}\\
	\caption{(a) Dispersion relation of the surface plasmon-magnon polariton. The light cone is bounded by $\omega =cq/\sqrt{\epsilon_4}$. $\epsilon_4=14,~\epsilon_3=2,~ E_F=0.8~\mathrm{eV}$. The parameters of YIG are used with $\epsilon_3 =10.8$, $M_s=0.175~\mathrm{T}$ \cite{Kuznetsov2019,notemo,KatsantonisSR2023}. (b) Resonant frequency of the surface plasmon-magnon polariton as a function of external field at different values of the Fermi energy in the hybrid structure shown in Fig. \ref{fig1}. $\theta=1.1 \theta_c$. The black line at $\omega=0$ is the solution to resonant condition \eqref{resonance_condition} in DE/2DEG/DE structure.}\label{fig2}
\end{figure}

The situation changes dramatically when medium 2 is a ferromagnet. By plugging $\delta_\mathrm{FM}$ into Eq. (\ref{resonance_condition}), we find that the resonant frequency should satisfy the equation
\begin{equation}\label{mpdispersion}
\sqrt{q^2-\frac{\omega^2 \epsilon_3}{c^2}} + \frac{q^2-k_2^2\nu}{q(1+\kappa + \nu)} + \frac{\mu_0\sigma_0 E_F}{\pi \hbar} =0.
\end{equation}
Firstly, we recover the frequency of surface magnon mode in the magnetostatic limit ($\omega \ll cq$) when $E_F=0$ as $\omega_r=\omega_h + \omega_m/2$ (blue and red dashed lines in Fig. \ref{fig2}(a)) \cite{DamonPR1960, SergaJPD2010}. As we go beyond this limit, the spectrum can be obtained by numerically solving Eq. (\ref{mpdispersion}), with the result shown in Fig. \ref{fig2}(a). Clearly, there is an overlap between the light cone and surface plasmon-magnon dispersion, suggesting the possibility to match both momentum and energy between the incident photons and hybrid plasmon-magnon modes and thus enabling the plasmon excitations. It is noteworthy that the resonant frequency can be well tuned in the GHz regime by the external field, as shown in Fig. \ref{fig2}(b).


{\it Reflection rate.}--Now we proceed to demonstrate that the surface plasmons and magnons can be simultaneously excited by shining a proper wave on the hybrid system. The excited surface plasmon will carry away electromagnetic energy and reduce the reflection rate of the system, which provides a feasible way to detect the excitation of surface plasmons in experiments.
Here we consider a $s$-polarized incident wave with electric field perpendicular to the incident plane $\mathbf{k}^{(i)}=(0,k_{y},k_{z})$ in medium 4, where $k_{y}=k_4\sin \theta$ and $k_{z}=k_4\cos \theta$, as shown in Fig. \ref{fig1}. To satisfy the Maxwell's equations, the magnetic and electric fields should read $\mathbf{H}_4^{(i/r)}=(0,H_{4,y}^{(i/r)},H_{4,z}^{(i/r)})e^{i(k_{y} y \pm k_{z}z)}$ and $\mathbf{E}_4^{(i/r)}=(E_{4,x}^{(i/r)},0,0)e^{i(k_{y} y \pm k_{z}z)}$ with $H_{4,y}^{(i/r)}=\pm E_{4,x}^{(i/r)} k_{z}/(\mu_0 \omega),~H_{4,z}^{(i/r)}=-E_{4,x}^{(i/r)} k_{y}/(\mu_0 \omega)$, where $i,r$ label the incident and reflected waves, respectively. Above a critical angle $\theta_c = \arcsin{\sqrt{\epsilon_3/\epsilon_4}}$ ($\epsilon_4>\epsilon_{3}$), the incident light induces evanescent waves in medium 3 \cite{Jackson1998EM}. The finite thickness of medium 3 allows for the coexistence of exponential increase and decay modes, {\em i.e.}
\begin{equation}
\begin{aligned}
\mathbf{H}_3 &=  (0,H_{3,y}^{(-)}, H_{3,z}^{(-)})e^{ik_{3,y}y-\kappa_3z}\\
&+(0,H_{3,y}^{(+)}, H_{3,z}^{(+)})e^{ik_{3,y}y+\kappa_3z},\\
\mathbf{E}_3 &= (E_{3,x}^{(-)}, 0,0)e^{ik_{3,y}y-\kappa_3z} + (E_{3,x}^{(+)}, 0,0)e^{ik_{3,y}y+\kappa_3z},
\end{aligned}
\end{equation}
where $H_{3,y}^{(\pm)} = \mp iE_{3,x}^{(\pm)} \kappa_3/(\mu_0 \omega)$ and $H_{3,x}^{(\pm)} = - iE_{3,x}^{(\pm)} k_{3,y}/(\mu_0 \omega)$.

\begin{figure}
	\centering
	\includegraphics[width=0.49\textwidth]{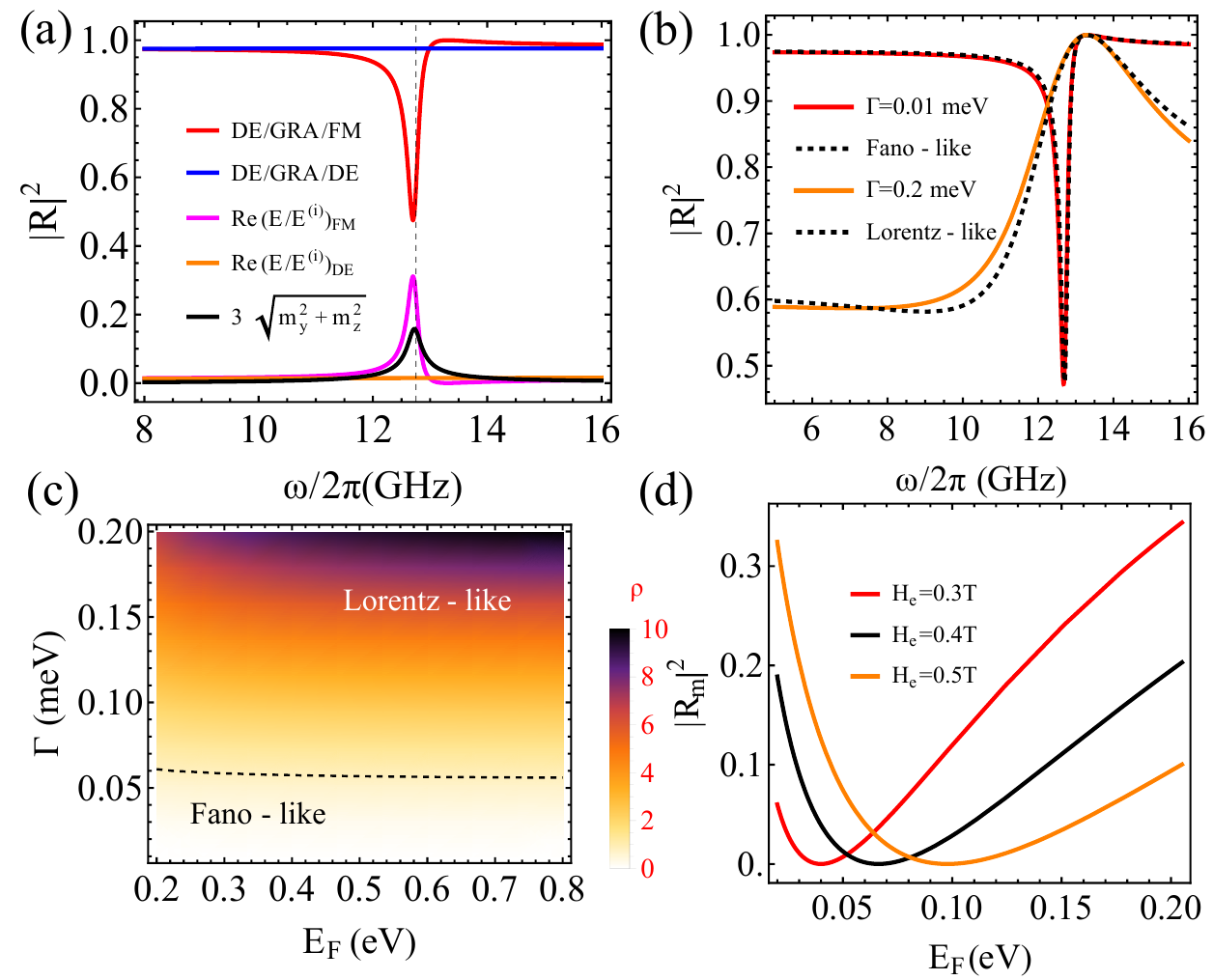}\\
	\caption{ (a) Reflection rate of the hybrid system, electric field strength at the FM surface, spin-wave excitation amplitude as a function of the incident wave frequency. $d=2.5~ \mathrm{\mu m},~H_e=0.3~\mathrm{T}, \alpha=10^{-4}, \Gamma=0.01~\mathrm{meV}, \theta=1.1 \theta_c, E_F=0.3~\mathrm{eV}$. (b) Fano-like and Lorentz-like reflection spectrum at small relaxation rate and large relaxation rate of carriers, respectively. The dashed lines are the results of analytical formula Eq. \eqref{RFanoLorentz}. (c) Density plot of the lineshape index $\rho$ in the $E_F-\Gamma$ plane. The lineshape is Fano-like for $\rho \ll 1$ and Lorentz-like for $\rho \gg 1$. The black dashed line is $\rho=1$. (d) The minimum reflection rate as a function of the Fermi energy at different external fields. $\Gamma=0.01~\mathrm{meV}$.}\label{fig3}
\end{figure}

Now the boundary conditions require the continuity of the tangential components of both electric and magnetic fields at interfaces of media 4-3 and 3-2, i.e.
\begin{subequations}
\begin{align}
&H_{4,y}^{(i)} + H_{4,y}^{(r)} = H_{3,y}^{(+)} + H_{3,y}^{(-)},\\
&E_{4,x}^{(i)} + E_{4,x}^{(r)} = E_{3,x}^{(+)} + E_{3,x}^{(-)},\\
&H_{3,y}^{(+)}e^{(\kappa_2+ \kappa_3)d} + H_{3,y}^{(-)}e^{(\kappa_2-\kappa_3) d}-\sigma E_{2,x}^{(-)}=H_{2,y}^{(-)},\\
&E_{3,x}^{(+)}e^{(\kappa_2+ \kappa_3) d} + E_{3,x}^{(-)}e^{(\kappa_2- \kappa_3) d} = E_{2,x}^{(-)}.
\end{align}
\end{subequations}
By expressing all the magnetic fields by their electric fields counterparts and solving the resulting linear equations, we can derive the reflection coefficient $R \equiv E_{4,x}^{(r)}/E_{4,x}^{(i)}$ as \cite{Supplement}
\begin{equation}
R=\frac{k_2^2 (k_z \sinh(\kappa_3d)-i\kappa_3\cosh(\kappa_3d))+\delta_p c_+}{k_2^2 (k_z \sinh(\kappa_3d)+i\kappa_3\cosh(\kappa_3d))+\delta_p c_-},
\end{equation}
where $c_\pm=(\mp\mu_0 \sigma \omega +k_z) \kappa_3 \cosh(\kappa_3d)\mp i(\kappa_3^2 \pm k_z \mu_0 \sigma \omega)\sinh(\kappa_3d)$.
For very thin dielectric medium 3 ($\kappa_3d \rightarrow 0$), the reflection coefficient is simplified as
\begin{equation}\label{R2lowd}
R=\frac{-ik_2^2 + \delta_p(k_z-\mu_0 \sigma \omega)}{ik_2^2 + \delta_p(k_z+\mu_0 \sigma \omega)}.
\end{equation}
This is the second key result of our work. Figure \ref{fig3}(a) shows the reflection rate $|R|^2$ as a function of the frequency of incident wave when $\theta =3 \theta_c$.
A sharp dip in the reflection rate appears at the resonant frequency (vertical dashed line), implying a resonant excitation of the surface plasmon-magnon polariton. As a comparison, the reflection rate is approximately one when the magnetic layer is replaced by a normal dielectric with the same permittivity $\epsilon_2$ (blue line), indicating very weak plasmon excitations. This comparison explicitly confirms that the magnetic layer releases the constraint to excite the TE surface plasmon. To understand the essential physics, we further plot the electric field in the 2DEG layer as well as the spin-wave amplitude as a function of the wave frequency in Fig. \ref{fig3}(a). When medium 2 is a magnetic layer, the spin-wave is maximally excited at resonance, which also significantly enhances the electric field in the 2DEG layer and thus strongly excites the surface plasmon mode. However, there is no enhancement of electric fields when medium 2 is a dielectric. Now it seems safe to conclude that the surface spin waves boost the surface plasmon excitations, which carry away significant amount of  electromagnetic energy and thus generate a considerable dip in the reflection spectrum.

{\it Lineshape of the reflection spectrum.}--- We further notice that the lineshape of the reflection spectrum near the resonance is asymmetric. Physically, this may be interpreted as an interference effect between the background continuum spectrum and a discrete mode \cite{FanoPR1961}. Here the continuous mode is the flat reflection spectrum without considering the magnetic properties of medium 2 (blue line in Fig. \ref{fig3}(a)) while the discrete mode is the hybrid surface plasmon-magnon mode. Specifically, we can expand the reflection rate \eqref{R2lowd} around the resonance frequency and derive that \cite{Supplement}
\begin{equation}\label{RFanoLorentz}
|R|^2 =A_0 \frac{(\omega - \omega_0 + \lambda \beta)^2 + \eta^2}{(\omega - \omega_0)^2 + \beta^2},
\end{equation}
where $\lambda$ is the well-known Fano parameter, $\omega_0=\omega_r - \Delta \omega$ is the modified resonance frequency, $\beta$ is the effective linewidth, and $\eta$ is the strength of the Lorentz contribution. In general, near the resonance position, we may characterize the relative weight of the Fano and Lorentz lineshapes by defining a lineshape index $\rho \equiv \eta/ (\Delta \omega + \lambda \beta)$ as \cite{Supplement}
\begin{equation}
\rho= \left |\frac{\pi k_z \Gamma^2 -\mu_0 \sigma_0 \omega E_F \Gamma + \pi k_z (\hbar \omega)^2 }{\mu_0 \sigma_0 E_F \hbar \omega^2} \right|.
\end{equation}

When the relaxation rate of carriers $\Gamma$ in 2DEG is very small, the ratio $\rho \approx \pi k_z \hbar/\mu_0 \sigma_0 E_F$  is much smaller than one for higher Fermi energy, then the reflection spectrum is Fano-like \cite{FanoPR1961}, as shown in Fig. \ref{fig3}(b) (red line).
When the relaxation rate $\Gamma$ is very high, the ratio becomes $\rho \approx  \pi k_z \Gamma^2/ (\mu_0 \sigma_0 E_F \hbar \omega^2)$. In this regime, the Lorentz contribution can be comparable and even dominate the Fano contribution (orange line in Fig. \ref{fig3}(b)). The complete phase diagram of $\rho$ in the $E_F-\Gamma$ plane is shown in Fig. \ref{fig3}(c).
It is noteworthy that the Fermi energy of the 2DEG can be tuned by electric gating \cite{CraciumNT2011}, which makes it possible to tune the lineshape and thus the minimum reflection rate of the hybrid system. Figure \ref{fig3}(d) shows that the minimum reflection rate $|R_m|^2$ can reach zero by appropriately tuning the Fermi energy and external fields. In this situation, all the energy of incident wave is converted to excite surface plasmons.

\begin{figure}
	\centering
	\includegraphics[width=0.4\textwidth]{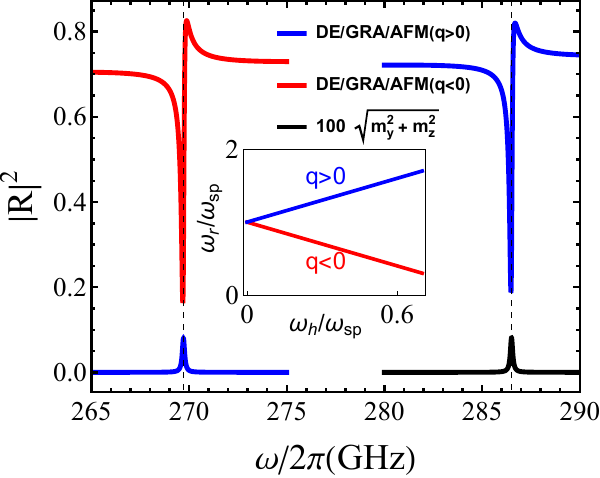}\\
	\caption{(a) Reflection rate and spin-wave excitation amplitude of an AFM as a function of incident frequency in an DE/GRA/AFM structure. Here, $\mathbf{m}$ is the magnetization order of an AFM. The inset shows the external field dependence of the frequency of plasmon-magnon mode. $\omega_\mathrm{sp}=\gamma \sqrt{H_\mathrm{an}(2H_\mathrm{ex}+H_\mathrm{an})}$. Parameters of $\mathrm{MnF}_2$ are used with exchange field $H_\mathrm{ex} = 55.6~\mathrm{T}$, anisotropy $H_\mathrm{an}=0.88~\mathrm{T}, M_s= 0.059~\mathrm{T}, \epsilon_2= 7.645$ \cite{JohnsonPR1959,SeehraJAP1984}, $\Gamma = 0.8~\mathrm{meV}$. Other parameters are the same as Fig. \ref{fig3}(a).}\label{fig4}
\end{figure}

{\it Extension to antiferromagnet.}--- The essential physics presented above can be extended to AFMs. As an example, we consider a two-sublattice AFM insulator with the easy axis and external field both aligned along the $x$-axis. Following a similar approach presented above, we can derive the reflection coefficient \cite{Supplement}. Figure \ref{fig4} shows the reflection rate as a function of the incident wave frequency. Unlike the ferromagnetic case, two distinguished dips appear in the sub-THz regime (red and blue lines) depending on the direction of in-plane momentum ($q$) of incident wave.
This difference is because there are two surface spin-wave modes in an AFM propagating in $\pm e_y$ directions respectively \cite{CamleyPRL1980}. The incident wave with $q>0$ ($q<0$) only excites the surface spin wave and plasmon propagating in the $+e_y$ ($-e_y$) directions. Therefore one may generate nonreciprocal surface plasmons \cite{Katsantonis2023} by properly choosing the wave frequency.


{\it Discussions and conclusions.}---In conclusion, we have shown that surface spin waves in both ferromagnets and antiferromagnets can boost the excitation of TE surface plasmons ranging from GHz to THz regime in 2D materials. The excitation condition is not constrained by 2D conductivity and is thus applicable to a wide class of 2D systems. The excitation of surface plasmons carries away electromagnetic energy and generates a local minimum in the reflection spectrum of the system. The dip structure is quite robust against conductivity variation of 2D layers caused by the adjacent magnetic material, the damping of magnetic systems, and the loss of dielectric layers caused by defects and disorder in real materials \cite{Supplement}. Our findings should open a novel and feasible hybrid platform to study the surface plasmons and further promote its application in designing spintronic and plasmonic devices. 



{\it Acknowledgments.}--- The work was supported by the National Key R$\&$D Program of China (2022YFA1402700) and the Dutch Research Council (NWO). H.Y.Y acknowledges the helpful discussions with Mathias Kl\"{a}ui, Rembert Duine, Alexander Mook, Pieter Gunnink, Artem Bondarenko, Mikhail Cherkasskii and Zhaoju Yang.

\bibliography{magnon_plasmon}

\begin{thebibliography}{44}
\bibitem [{not()}]{notemo}%
  \BibitemOpen
  \href@noop {} {}\bibinfo {note} {Here we neglect the modification of electric
  permittivity by magneto-optical effects, which may only become relevant in
  the optical reigme \cite{KatsantonisSR2023}.}\BibitemShut {Stop}%
\end{thebibliography}
\end{document}